# Report

# LavAtmos: An open-source chemical equilibrium vaporization code for lava worlds

Christiaan P. A. van BUCHEM 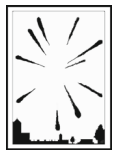[1]*, Yamila MIGUEL [1,2], Mantas ZILINSKAS [1], and Wim van WESTRENEN [3]

[1]Leiden Observatory, Leiden University, Leiden, The Netherlands
[2]SRON Netherlands Institute for Space Research, Leiden, The Netherlands
[3]Faculty of Science, Earth Sciences, Vrije Universiteit Amsterdam, Amsterdam, The Netherlands
*Correspondence
Christiaan P. A. van Buchem, Leiden Observatory, Leiden University, Leiden, The Netherlands.
Email: vbuchem@strw.leidenuniv.nl

*Abstract*–To date, over 500 short-period rocky planets with equilibrium temperatures above 1500 K have been discovered. Such planets are expected to support magma oceans, providing a direct interface between the interior and the atmosphere. This provides a unique opportunity to gain insight into their interior compositions through atmospheric observations. A key process in doing such work is the vapor outgassing from the lava surface. LavAtmos is an open-source code that calculates the equilibrium chemical composition of vapor above a dry melt for a given composition and temperature. Results show that the produced output is in good agreement with the partial pressures obtained from experimental laboratory data as well as with other similar codes from literature. LavAtmos allows for the modeling of vaporization of a wide range of different mantle compositions of hot rocky exoplanets. In combination with atmospheric chemistry codes, this enables the characterization of interior compositions through atmospheric signatures.

## INTRODUCTION

With an ever-growing catalog of newly discovered exoplanets, we have moved from the discovery phase well into the characterization phase. An emerging category of specific interest is that of the so-called hot rocky exoplanets.

These planets are exposed to extreme stellar irradiation that leads to surface temperatures hot enough to prevent the planet from cooling and creating a crust (Boukaré et al., 2022; Henning et al., 2018), exposing the silicate mantle directly to the atmosphere. Furthermore, since the atmosphere is a direct product of the outgassing from the magma ocean and is in equilibrium with it, the atmospheric compositions of these planets are directly influenced by the interior composition (Dorn & Lichtenberg, 2021; Ito et al., 2015; Kite et al., 2016, 2020; Miguel et al., 2011; Nguyen et al., 2020). This provides a unique opportunity to derive interior properties from atmospheric observations. In addition, there is growing evidence that planetesimals, as well as the rocky planets and moons that form from their accretion, were covered by magma early in their evolution (Elkins-Tanton, 2012; Greenwood et al., 2005; Hin et al., 2017; Norris & Wood, 2017; Schaefer & Elkins-Tanton, 2018). Therefore, constraints on the atmospheric products of interior–atmosphere interactions on hot rocky exoplanets might also provide us with a window to the conditions in the early solar system and early Earth (Hirschmann, 2012).

Hot rocky planets have been the targets of several observing programs throughout the past several years on a range of different ground- and space-based telescopes (Deibert et al., 2021; Demory et al., 2011; Esteves et al., 2017; Keles et al., 2022). Conclusions drawn from these observations remain uncertain and have yet to give definitive proof of an atmosphere. However, some tentative evidence has been given for K2-141 b (Zieba et al., 2022) and 55 Cnc e (Angelo & Hu, 2017; Demory et al., 2016; Zilinskas et al., 2020, 2021), and the advent of the new generation of telescopes, such as JWST and Ariel, may allow for the characterization of the chemical composition of hot rocky-exoplanet atmospheres in the near future (Ito et al., 2021; Zilinskas et al., 2022).

To know what to look for in such observations and to interpret the data once it arrives, accurate atmospheric models are required. For atmospheres on hot rocky exoplanets, this involves modeling the degassing from lava at the surface of the planet. Since we yet have to gain a good understanding of the possible types of rocky-exoplanet compositions, we do not yet know what kind of compositions we should expect for these melts. Recent work (Brugman et al., 2021; Putirka et al., 2021; Putirka & Rarick, 2019) indicates that we should expect a wide range of different possible silicate compositions. Hence, open-source vaporization codes that can work with a wide range of compositions are necessary to enable modeling potential atmospheres as our understanding of hot rocky planets develops.

To date, a limited number of codes have been used to calculate the chemical composition of vapors degassing from lava at a given temperature and the composition of an atmosphere in equilibrium with lava of a given temperature. The MAGMA code (Fegley & Cameron, 1987) was written to study the fractional vaporization of Mercury. The same code was used for the study of other solar system bodies (Schaefer & Fegley, 2007; Schaefer & Fegley Jr., 2004) and exoplanets (Kite et al., 2016; Miguel et al., 2011; Schaefer & Fegley, 2009; Schaefer et al., 2012; Visscher & Fegley, 2013). This code makes use of the Ideal Mixing of Complex Components (IMCC) model, developed by Hastie and co-authors (Hastie & Bonnell, 1985, 1986; Hastie et al., 1982), to calculate the activity of the oxide components in the melt.

In more recent years, the MELTS code (Ghiorso & Sack, 1995) has increased use for modeling the thermodynamics for outgassing codes (Ito et al., 2015, 2021; Jäggi et al., 2021; Wolf et al., 2023). Wolf and co-authors have developed the code named VapoRock which has been used to model the early atmosphere of Mercury (Jäggi et al., 2021) and to explore how relative abundances of SiO and SiO$_2$ could be used to infer the O$_2$ fugacity of a volatile-depleted mantle (Wolf et al., 2023). The main difference between VapoRock and LavAtmos is the manner in which the O$_2$ partial pressure is determined. In the discussion ("Discussion" Section), we include a more in-depth comparison of the two codes. Other approaches that calculate the condensate compositions from an initial gas composition, as opposed to calculating vaporization reactions from an existing melt reservoir, have also been developed in the literature (Herbort et al., 2020, 2022).

In this paper, we present a new open-source code, which we named LavAtmos, that calculates the equilibrium composition of a vapor above a melt of a given composition, at a given temperature and at a given melt–vapor interface pressure. As shown in the graphical table of contents, a general overview of the workflow of the code is presented. Just as the abovementioned previous works (Ito et al., 2015, 2021; Wolf et al., 2023), we use MELTS to calculate the oxide component properties of a melt. These properties are then combined with thermochemical data available in the JANAF tables (Chase, 1998) to perform gas–melt equilibrium calculations. The oxygen fugacity ($f$O$_2$) is derived from the law of mass action, similarly to the approach used for thermodynamic calculations for pure silica and alumina (Krieger, 1965a, 1965b) and for the MAGMA code (Fegley & Cameron, 1987). LavAtmos currently takes 9 oxide species into account (SiO$_2$, MgO, Al$_2$O$_3$, TiO$_2$, Fe$_2$O$_3$, FeO, CaO, Na$_2$O, and K$_2$O), 31 different vapor species with corresponding vaporization reactions (shown in Table 1) and is suitable for calculations between 1500 and 4000 K. LavAtmos is written in Python for ease of use and integration with the MELTS Python wrapper named Thermoengine.[1] It is released as an open-source code under the GNU General Public License version 3.[2] LavAtmos is available on https://github.com/cvbuchem/LavAtmos.

In this paper, we provide an in-depth look at the methods used in "Methodology" Section. We compare its performance to laboratory data and the results calculated by other similar codes where those are available in the public domain in "Validation" Section. Finally, we discuss assumptions made in the method, the advantages and limitations of the code, and highlight a set of potential applications in "Discussion" Section, rounding off with a conclusion in "Conclusion" Section.

## METHODOLOGY

In this section, we cover how the partial pressures of included species are calculated. Consider the generalized form of a vaporization reaction of a liquid oxide $j$, gaseous O$_2$, and the resulting vapor species $i$:

$$c_{ij}X_{x_j}O_{y_j}(\mathrm{l}) + d_{ij}O_2(\mathrm{g}) \Leftrightarrow X_{x_jc_{ij}}O_{y_jc_{ij}+2d_{ij}}(\mathrm{g}) \qquad (1)$$

$X$ is the cation of the species, $x_j$ the number of cation atoms, $y_j$ the number of oxygen atoms, and $c_{ij}$ and $d_{ij}$ are the stoichiometric coefficients for this reaction. The gaseous species (atmosphere) are indicated using (g) and the liquid species (melt) using (l). As an example, the

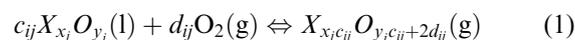

TABLE 1. Overview of the 31 vaporization reactions included in LavAtmos.

| End-member | # | Reactants | | Vapor |
|---|---|---|---|---|
| – | 1 | $1/2O_2(g)$ | ↔ | $O(g)$ |
| $SiO_2$ | 2 | $SiO_2(l) - O_2(g)$ | ↔ | $Si(g)$ |
| | 3 | $2SiO_2(l) - 2O_2(g)$ | ↔ | $Si_2(g)$ |
| | 4 | $3SiO_2(l) - 3O_2(g)$ | ↔ | $Si_3(g)$ |
| | 5 | $SiO_2(l) - 1/2O_2(g)$ | ↔ | $SiO(g)$ |
| | 6 | $SiO_2(l)$ | ↔ | $SiO_2(g)$ |
| $Al_2O_3$ | 7 | $1/2Al_2O_3(l) - 3/4O_2(g)$ | ↔ | $Al(g)$ |
| | 8 | $Al_2O_3(l) - 3/2O_2(g)$ | ↔ | $Al_2(g)$ |
| | 9 | $1/2Al_2O_3(l) - 1/4O_2(g)$ | ↔ | $AlO(g)$ |
| | 10 | $Al_2O_3(l) - O_2(g)$ | ↔ | $Al_2O(g)$ |
| | 11 | $1/2Al_2O_3(l) - 1/2O_2(g)$ | ↔ | $AlO_2(g)$ |
| | 12 | $Al_2O_3(l) - 1/2O_2(g)$ | ↔ | $Al_2O_2(g)$ |
| $TiO_2$ | 13 | $TiO_2(l) - O_2(g)$ | ↔ | $Ti(g)$ |
| | 14 | $TiO_2(l) - 1/2O_2(g)$ | ↔ | $TiO(g)$ |
| | 15 | $TiO_2(l)$ | ↔ | $TiO_2(g)$ |
| $Fe_2O_3$ | 16 | $1/2Fe_2O_3(l) - 3/4O_2(g)$ | ↔ | $Fe(g)$ |
| | 17 | $1/2Fe_2O_3(l) - 1/4O_2(g)$ | ↔ | $FeO(g)$ |
| $Fe_2SiO_4$ | 18 | $1/2Fe_2SiO_4(l) - 1/2O_2(g) - 1/2SiO_2(l)$ | ↔ | $Fe(g)$ |
| | 19 | $1/2Fe_2SiO_4(l) - 1/2SiO_2(l)$ | ↔ | $FeO(g)$ |
| $Mg_2SiO_4$ | 20 | $1/2Mg_2SiO_4(l) - 1/2O_2(g) - 1/2SiO_2(l)$ | ↔ | $Mg(g)$ |
| | 21 | $Mg_2SiO_4(l) - O_2(g) - SiO_2(l)$ | ↔ | $Mg_2(g)$ |
| | 22 | $1/2Mg_2SiO_4(l) - 1/2SiO_2(l)$ | ↔ | $MgO(g)$ |
| $CaSiO_3$ | 23 | $CaSiO_3(l) - 1/2O_2(g) - SiO_2(l)$ | ↔ | $Ca(g)$ |
| | 24 | $2CaSiO_3(l) - O_2(g) - 2SiO_2(l)$ | ↔ | $Ca_2(g)$ |
| | 25 | $CaSiO_3(l) - SiO_2(l)$ | ↔ | $CaO(g)$ |
| $Na_2SiO_3$ | 26 | $1/2Na_2SiO_3(l) - 1/4O_2(g) - 1/2SiO_2(l)$ | ↔ | $Na(g)$ |
| | 27 | $Na_2SiO_3(l) - 1/2O_2(g) - SiO_2(l)$ | ↔ | $Na_2(g)$ |
| | 28 | $1/2Na_2SiO_3(l) + 1/4O_2(g) - 1/2SiO_2(l)$ | ↔ | $NaO(g)$ |
| $KAlSiO_4$ | 29 | $KAlSiO_4(l) - 1/4O_2(g) - 1/2Al_2O_3(l) - SiO_2(l)$ | ↔ | $K(g)$ |
| | 30 | $2KAlSiO_4(l) - 1/2O_2(g) - Al_2O_3(l) - 2SiO_2(l)$ | ↔ | $K_2(g)$ |
| | 31 | $KAlSiO_4(l) + 1/4O_2(g) - 1/2Al_2O_3(l) - SiO_2(l)$ | ↔ | $KO(g)$ |

*Note*: Reaction 1 is the gas to gas reaction of $O_2$ to O. The rest of the reactions (2–31) describes the vaporization reactions. Each vapor species has a unique reaction that includes a MELTS end-member species, $O_2$, and (if necessary) residue liquid metal oxides.

vaporization reaction of liquid $SiO_2$ to form gaseous SiO can be written as:

$$SiO_2(l) - \frac{1}{2}O_2(g) \Leftrightarrow SiO(g) \qquad (2)$$

Assuming that the reaction is in equilibrium, the partial pressure of the vapor species $i$ can be calculated by adhering to the law of mass action as follows:

$$P_{ij} = K_{r_{ij}} a_j^{c_{ij}} P_{O_2}^{d_{ij}} \qquad (3)$$

where $P_{ij}$ is the partial pressure of vapor species $i$ as formed from liquid species $j$, $K_{r_{ij}}$ is the chemical equilibrium constant of the reaction, $a_j$ is the activity of liquid oxide $j$, and $P_{O_2}$ is the partial pressure of $O_2$ (also known as the oxygen fugacity $fO_2$). Some more elaboration on the derivation is shown in the appendix ("Deriving the Partial Pressure Equation" Section). For

the example reaction shown in Equation (2), the partial pressure of SiO can be determined using:

$$P_{SiO} = K_r a_{SiO_2} P_{O_2}^{-1/2} \qquad (4)$$

The variables that must be known to calculate the partial pressure of a vapor species are the stoichiometric coefficients $c_i$ and $d_i$, the chemical equilibrium constant of reaction $K_{r_i}$ for the vapor species $i$, the chemical activity $a_j$ of the oxide $j$ involved in the reaction, and the oxygen partial pressure $P_{O_2}$. Stoichiometric coefficients are determined by writing out balanced reaction equations (see Table 1). The chemical equilibrium constant of each reaction $K_{r_{ij}}$ is determined using the data available in the JANAF tables (Chase, 1998).

The activity of the oxides in the melt is determined using the MELTS code. Developed over the course of the past two and a half decades, MELTS has been consistently updated and expanded (Asimow & Ghiorso, 1998; Ghiorso & Gualda, 2015; Ghiorso et al., 2002; Ghiorso & Sack, 1995; Gualda et al., 2012). It performs internally consistent modeling of liquid–solid equilibria in magmatic systems at elevated temperatures and pressures. For a given temperature, pressure, and composition (in terms of oxide weight percentages, see Table 1 for example compositions), MELTS is able to calculate the thermochemical properties of the end-member component species that it accounts for in each included phase (Ghiorso & Sack, 1995). In the current version of LavAtmos, only the liquid phase is used. Hence, as of yet, partial melting is not taken into account. The properties, specifically the activities, of these end-member components are used by LavAtmos to calculate the energetics of the vaporization reactions of the oxides in the melt. The end-member species included in our MELTS calculations are $SiO_2$, $Al_2O_3$, $TiO_2$, $Fe_2O_3$, $Fe_2SiO_4$, $Mg_2SiO_4$, $CaSiO_3$, $Na_2SiO_3$, and $KAlSiO_4$.[3] Due to the fact that some of these species include more than one cation, we write vaporization reactions that produce a vapor species for one of the cations and liquid species for the others, similarly to the approach taken in work done on condensates from impacts between silicate-rich bodies (Fedkin et al., 2006). Using $KAlSiO_4$ as example, the equilibrium vaporization reaction may be written as:

$$KAlSiO_4(l) - \tfrac{1}{4}O_2(g) - \tfrac{1}{2}Al_2O_3(l) - SiO_2(l) \Leftrightarrow K(g) \quad (5)$$

Therefore, the partial pressure for $K$ is calculated using:

$$P_K = K_r a_{KAlSiO_4} P_{O_2}^{-1/4} a_{Al_2O_3}^{-1/4} a_{SiO_2}^{-1} \quad (6)$$

Since $Al_2O_3$ and $SiO_2$ are also end-member species, their activities are known. In the cases where a gas species has more than one reaction producing it, such as for Fe (reactions 16 and 18 in Table 1) and FeO (reactions 17 and 19 in Table 1), the total partial pressure of the species is equal to the sum of the partial pressures calculated for each reaction.

MELTS also requires a pressure as input. For the results shown in this paper, we assume a surface pressure of 0.1 bar. When testing the results for a range of pressure from $10^{-4}$ up to 100 bar, there were no significant differences in the resulting partial pressures. This pressure range includes the atmospheric pressures expected at the lava–atmosphere interface on hot rocky exoplanets (Zilinskas et al., 2022). We assume the silicate compositions are fully molten and for this reason only include the liquid phase in the MELTS calculations. The Python wrapper named Thermoengine is used to interact with the MELTS code (written in C). At the moment of submission of this manuscript, LavAtmos works with MELTS version 1.0.2. This version includes rhyolites-MELTS (Gualda et al., 2012) allowing for the modeling of hydrous silicic systems which may become relevant in future iterations of LavAtmos.

Looking back at Equation (3), we still need to determine the $O_2$ partial pressure ($P_{O_2}$). For this, we assume that the metal to oxygen ratio in the vapor is the same as the stoichiometries of the melt oxides, as is also done in the vapor calculations done by Krieger (1965a, 1965b). This gives us a mass balance equation in the form of:

$$n_{O_2} = -\sum_i \sum_j d_{ij} n_{ij} \quad (7)$$

The definition of partial pressure ($P_{ij} = \tfrac{n_{ij} P}{n}$ for vapor species $i$ produced from liquid species $j$) can be applied to rewrite Equation (7) as:

$$P_{O_2} = -\sum_i \sum_j d_{ij} P_{ij} \quad (8)$$

It is here that we see the importance of tracking the partial pressure of a vapor species $i$ released from a specific reaction with a liquid oxide $j$. For vapor species that have more than one reaction producing them, such as the aforementioned Fe and FeO, it is necessary to couple the corresponding partial pressures of released Fe/FeO ($P_{ij}$) to the right stoichiometry ($d_{ij}$). Substituting Equation (3) and setting one side to 0 gives:

$$0 = P_{O_2} + \sum_i \sum_j d_{ij} K_{r_{ij}} a_j^{c_{ij}} P_{O_2}^{d_{ij}} \quad (9)$$

Finding the $P_{O_2}$, for which Equation (9) holds, allows one to calculate the partial pressures of all other vapor species (using Equation (3)) while ensuring that the mass action law is being upheld. In Figure 1, the behavior of the right side of Equation (9) as a function of $P_{O_2}$ is shown for different melt temperatures. The higher the melt temperature, the greater the $P_{O_2}$ value at which Equation (9) holds true (where the y-axis is zero in Figure 1). If a $P_{O_2}$ value is used that corresponds to a (non-zero) negative value on the y-axis at a given temperature, this implies that not all of the $O_2$ released during vaporization reactions is included in the vapor. If a $P_{O_2}$ value is used that corresponds to a (non-zero) positive value on the y-axis at a given temperature, this implies that extra oxygen would have to be added to the system to reach the calculated equilibrium.

In LavAtmos, Equation (9) is solved for $P_{O_2}$ using the Scipy optimization function fsolve.[4] A vaporization reaction is defined for each vapor species included in LavAtmos (as shown in Table 1).

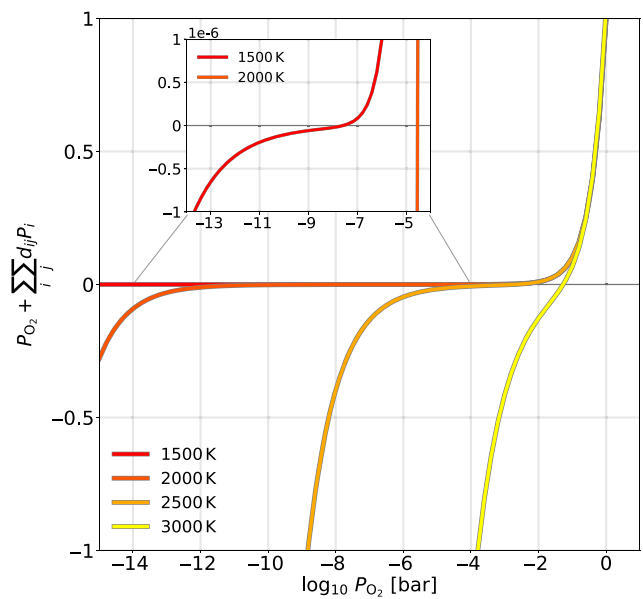

FIGURE 1. Behavior of the mass law function as a function of $P_{O_2}$ at different temperatures. LavAtmos determines the $P_{O_2}$ value by solving the mass law function shown in 9. Here we see the behavior of the function for different $P_{O_2}$ values and at different temperatures. An inset is used to show the behavior of the function at a melt temperature of 1500 K due to the difference in scale within which the changes occur relative to higher temperatures. (Color figure can be viewed at wileyonlinelibrary.com.)

## VALIDATION

In this section, LavAtmos results are compared to both experimental data and the results of other similar codes. In Figure 2, we compare the output of the LavAtmos code to that of the data gathered by Hastie and co-authors (Hastie & Bonnell, 1985; Hastie et al., 1982), with output from MAGMA (Schaefer & Fegley Jr., 2004), and output from VapoRock (Wolf et al., 2023) where available. Note that for the laboratory data, no uncertainties were provided by the authors.

In the panels on the left side of the figure (a, c, and e), the partial pressures of $O_2$, K, and Na above a synthetic (SRM) glass are shown. The bulk composition of this glass is given in Table 2. For $O_2$, we see that the results of both MAGMA and LavAtmos align well with the laboratory data. VapoRock is not compared in the $O_2$ plots due to it requiring $fO_2$ as an input (more on this below), for the calculation of the partial pressures of the other species VapoRock used the values given by Hastie and co-authors as input. For K, the partial pressure predicted by LavAtmos and VapoRock differ by about a factor 2, with VapoRock lying closer to the experimental data. The partial pressures predicted by MAGMA lie about one order of magnitude below that of VapoRock. For Na, it appears that MAGMA output falls closer to the laboratory data while LavAtmos and VapoRock slightly underestimate the Na partial pressures.

For Illite (see Table 2 for its bulk composition), on the right side of Figure 2 (b, d, and f), we see that the predicted partial pressure for $O_2$ is in agreement with the data at temperatures of 1750 K and above. We also see a much better agreement between the LavAtmos output and the data for K than we did for SRM glass, with VapoRock giving almost identical values while MAGMA tends to underestimate the K partial pressure by about a factor 2 at lower temperatures ($\sim 1550$ K) and about an order of magnitude at higher temperatures (above 2000 K). The partial pressure for SiO predicted by LavAtmos and VapoRock is very similar with both underestimating the experimental data by about a factor 2.

Overall we find good agreement between the calculated partial pressures and the experimental data. We do recognize that this is a limited amount of data to which we are able to compare our output and that the lack of experimental uncertainties makes it difficult to quantify the results. The difference in the values predicted for the K partial pressures by MAGMA and LavAtmos is due to the difference in the thermodynamic models used by the two codes. As explained in previous work (Ito et al., 2015) where similar discrepancies with the MAGMA code were found in their MELTS-based calculations, the IMCC model (used by MAGMA to calculate activity values) is calibrated by experimental results on synthetic high-$K_2O$ melts. This differs from MELTS, which is calibrated on experimental data drawn from natural, low$K_2O$ melts. This could also help explain why the K partial pressures predicted by LavAtmos appear to be closer to the experimental data for Illite (a naturally occurring composition) than for the SRM glass (synthetic composition). Another reason for the difference in partial pressures could be the much higher percentage of CaO and $Na_2O$ in synthetic melts than in natural compositions. This would also explain why VapoRock is showing similar behavior to that of LavAtmos. The difference between the values calculated by LavAtmos and VapoRock is likely due to the different treatments of $fO_2$. We will elaborate further upon this below.

To illustrate the use of LavAtmos and compare with other codes, we modeled the vapor output for a bulk silicate earth (BSE) composition (see Table 2) at temperatures between 1500 and 4000 K. The resulting partial pressures and mole fractions are shown in Figure 3. The dominant species up to about 2700 K is Na vapor, at which point SiO vapor takes over as the dominant species. $O_2$ and O have a strong presence as well throughout the majority of this temperature range. The general trend is that the more volatile species, such as Na, K, Fe, and Mg make up a larger fraction of the total

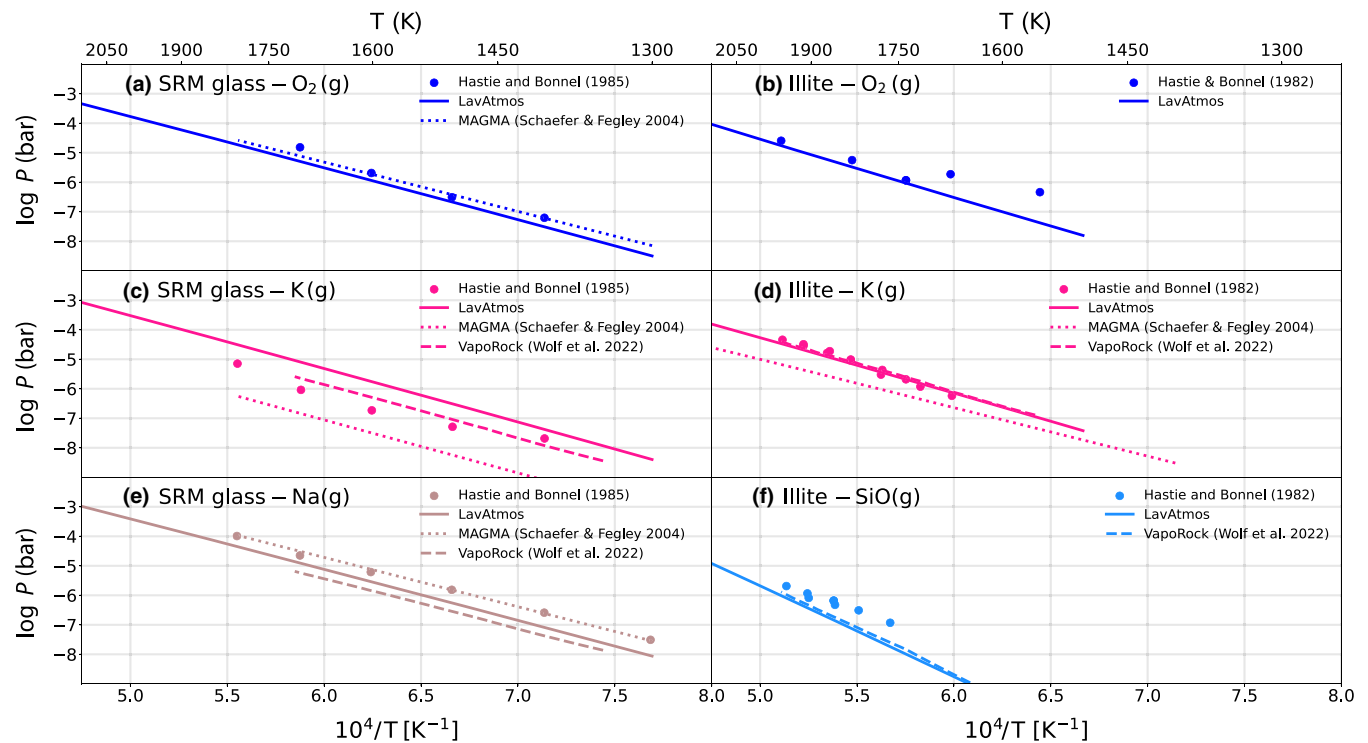

FIGURE 2. Comparison between data from Hastie and co-authors and code output. Each plot shows the partial pressure of a vapor species as experimentally measured by Hastie et al. (1982) and Hastie and Bonnell (1985) (dots), calculated using LavAtmos (solid line), calculated using MAGMA (dotted line) (Schaefer & Fegley Jr., 2004), and as calculated using VapoRock (dashed line) (Wolf et al., 2023). Plots (a, c and e) show the results for SRM glass, while plots (b, d and f) show the results for illite. The compositions of the melts are shown in Table 2. Not all species had available data from all codes; hence, some of the codes are missing in some of the plots. (Color figure can be viewed at wileyonlinelibrary.com.)

TABLE 2. Melt compositions used for Figures 2–6.

| Oxide wt% | SRM glass | Illite | BSE | CB | EH4 | NSP source | NSP lava |
|---|---|---|---|---|---|---|---|
| SiO$_2$ | 71.390 | 60.200 | 45.400 | 50.700 | 62.730 | 53.670 | 58.700 |
| MgO | 0.270 | 2.100 | 36.760 | 36.900 | 30.240 | 36.890 | 13.900 |
| Al$_2$O$_3$ | 2.780 | 26.000 | 4.480 | 4.600 | 2.580 | 4.750 | 13.800 |
| TiO$_2$ | – | – | 0.210 | – | – | – | – |
| Fe$_2$O$_3$ | 0.040 | 4.400 | – | – | – | – | – |
| FeO | – | – | 8.100 | 3.500 | – | 0.020 | 0.040 |
| CaO | 10.750 | – | 3.650 | 3.300 | 1.990 | 2.260 | 5.810 |
| Na$_2$O | 12.750 | 0.200 | 0.349 | 0.190 | 1.710 | 1.970 | 7.000 |
| K$_2$O | 2.020 | 7.400 | 0.031 | 0.050 | 0.200 | 0.050 | 0.200 |
| Total | 100.00 | 100.300 | 98.980 | 99.240 | 99.450 | 99.610 | 99.450 |

*Note*: The compositions given are for SRM glass (Hastie & Bonnell, 1985), Illite (Hastie et al., 1982), bulk silicate earth (BSE) (Palme & O'Neill, 2003), a CB chondrite chondrule with bulk Na and K mass balanced for chondrules to fit bulk meteorite iron–silicate ratio (Weisberg et al., 1990, 2001), an analog to enstatite chondrites (EH4) (Wiik, 1956), and lava Northern Smooth Planes (NSP) for both source and lava (Namur et al., 2016; Nittler & Weider, 2019).

gas pressure at lower temperatures than they do at higher temperatures. The opposite is true for the less volatile species, such as Si, Al, and Ti. In Figure 4, we can see a comparison of a selection of the calculated mole fractions for a BSE composition (Table 2) as calculated by LavAtmos, Ito and co-authors (Ito et al., 2015), and MAGMA (Visscher & Fegley, 2013). The BSE composition used for the Ito and co-authors results differs slightly (a few decimal percentages) from the BSE composition used for the other calculations. Tests using

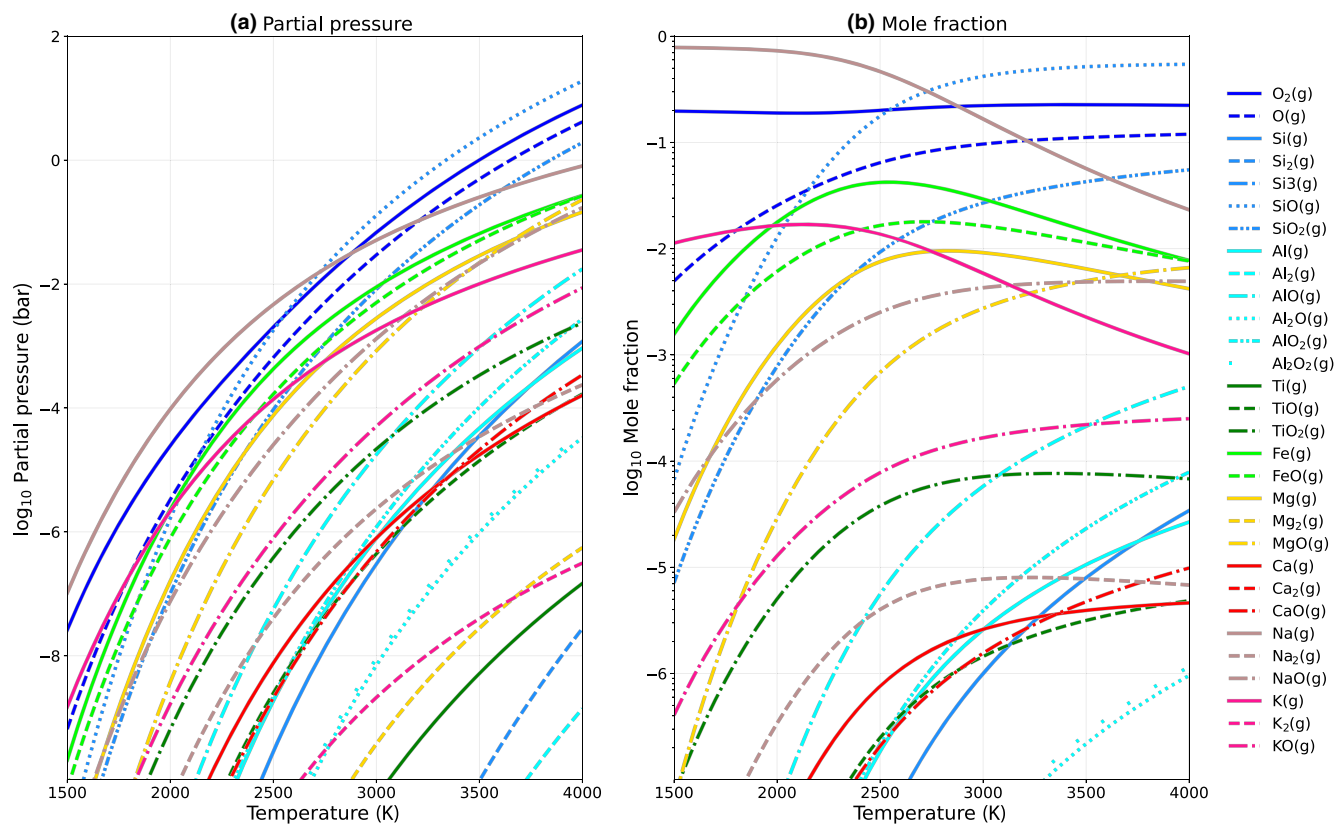

**FIGURE 3.** LavAtmos output for bulk silicate earth (BSE) composition between 1500 and 4000 K. The left panel (a) shows the partial pressures and the right panel (b) shows the mole fractions (partial pressure divided by the total pressure) of the included vapor species. These calculations were done for a BSE composition (see Table 2). A surface pressure of 0.01 bar was assumed. (Color figure can be viewed at wileyonlinelibrary.com.)

LavAtmos however show that the difference is so small that the effects are negligible. Due to lack of details on Ito and co-authors' (Ito et al., 2015) gas–melt equilibrium calculations, we are not able to comment on differences in methodology. In Figure 5, we also show the corresponding total pressures. The partial pressures calculated by Ito and co-authors (Figure 4a) are very close to the partial pressures calculated by LavAtmos. This is likely due to the fact that both codes make use of MELTS (Ghiorso & Sack, 1995) to calculate the thermodynamic properties of the melt. In addition to potential differences in the approach, our calculations include a wider range of vapor species. When comparing LavAtmos to the MAGMA code (Fegley & Cameron, 1987) data in recently published work (Visscher & Fegley, 2013) (Figure 4c), the most significant difference is between the calculated K partial pressures, as we also saw in Figure 2. As explained earlier in this section, this is due to the difference in calibration of the thermodynamic models used for the modeling of the liquid oxide activities (MELTS for LavAtmos and IMCC for MAGMA). The total vapor pressures calculated for each code (Figure 5) show similar results. At lower temperatures, it appears that LavAtmos and Ito and co-authors predict a lower total pressure than MAGMA. For higher temperatures, MAGMA predicts a lower total pressure compared to the other two codes. Note that the mole fractions of the vapor species are calculated by normalizing the partial pressures by the total pressure calculated by the codes. Hence, if two codes have the same mole fractions as output for certain vapor species, this may not translate to the same partial pressure values depending on whether the calculated total pressures are the same or not.

In Figure 6, we compare the output of LavAtmos (solid lines) with that of VapoRock (dashed lines) for four different compositions (see Table 2) as published in the work by Jäggi et al. (2021). VapoRock is a gas–melt equilibrium code that is also based on the thermodynamics calculated by MELTS. One of the ways in which this is apparent is the similar values calculated for the K partial pressures (pink) for all compositions. The greatest difference between the results of the two codes is seen in the values calculated for oxygen partial pressure. This can be attributed to the main difference in the approach taken by VapoRock and LavAtmos: the manner in which the oxygen fugacity of the system is determined.

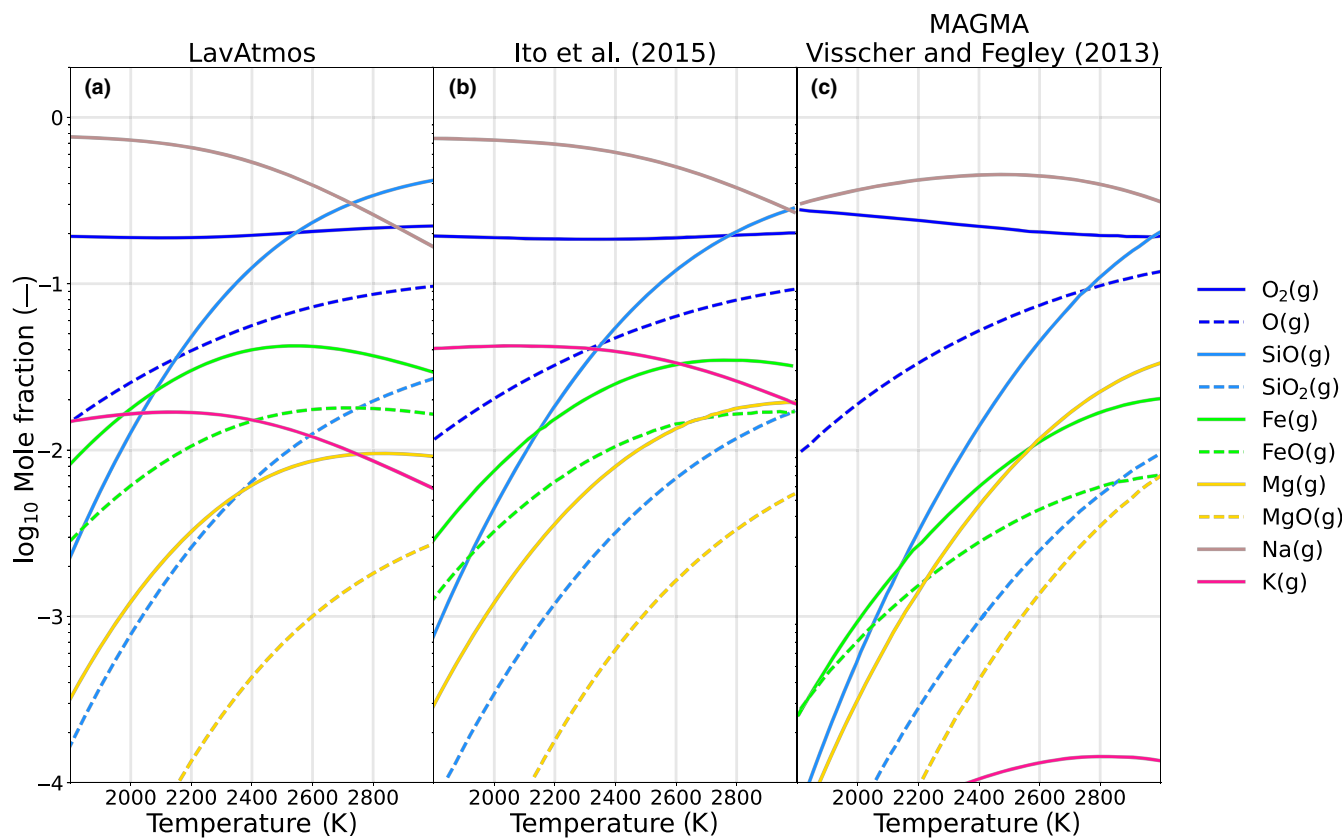

FIGURE 4. Comparison of mole fractions above a bulk silicate earth (BSE) melt as calculated by different codes. A selection of vapor species was made based on the data available in other works. The results shown are from LavAtmos (a), Ito et al. (2015) (b), and MAGMA (Visscher & Fegley, 2013) (c). (Color figure can be viewed at wileyonlinelibrary.com.)

For VapoRock, the oxygen fugacity of the melt is one of the user inputs (alongside temperature, pressure, and melt composition). For the results shown in Figure 6, an oxygen fugacity of IW-1 was used (Jäggi et al., 2021) (1 log unit less than the oxygen fugacity of the known ironwustite buffer). This provides the user the option to fix the oxygen fugacity value to a certain value, or to test the influence of different oxygen fugacity values on the outgassing of the other species.

LavAtmos, however, does not take oxygen fugacity as an input and instead calculates it internally through the dual constraints of mass action and mass balance, similarly to MAGMA (Fegley & Cameron, 1987; Schaefer & Fegley Jr., 2004), as shown in "Methodology" Section (Equations (3) and (7)). This allows the user to calculate a value for the oxygen fugacity based on the assumed congruent vaporization of the oxide species in the melt. As shown in Figure 2, this method is able to reproduce the oxygen fugacity values seen in experimental data. The same has been shown repeatedly for MAGMA (Fegley & Cameron, 1987; Schaefer & Fegley Jr., 2004; Sossi & Fegley, 2018). VapoRock makes use of the thermodynamic data from the work done by Lamoreaux and co-authors (Lamoreaux & Hildenbrand, 1984; Lamoreaux et al., 1987) and from the JANAF tables (Chase, 1998), while LavAtmos sources its thermodynamic data from the JANAF tables only. Lamoreaux and co-authors found that the thermodynamic values that they found were in good agreement with those found in the JANAF database. Hence, we expect that this difference in approach does not have a significant impact on the results. VapoRock includes degassing of chromium species (Cr), while LavAtmos does not. We expect that future iterations of LavAtmos will also include this species.

## DISCUSSION

LavAtmos assumes that thermochemical equilibrium is reached. Hence, whenever applying this code to a system one should keep in mind whether or not the time scales of the system allow for chemical equilibrium to be reached (at least on a local level). Due to the high temperatures (>1500 K) for which the code is meant, it can generally be safely assumed that an equilibrium state

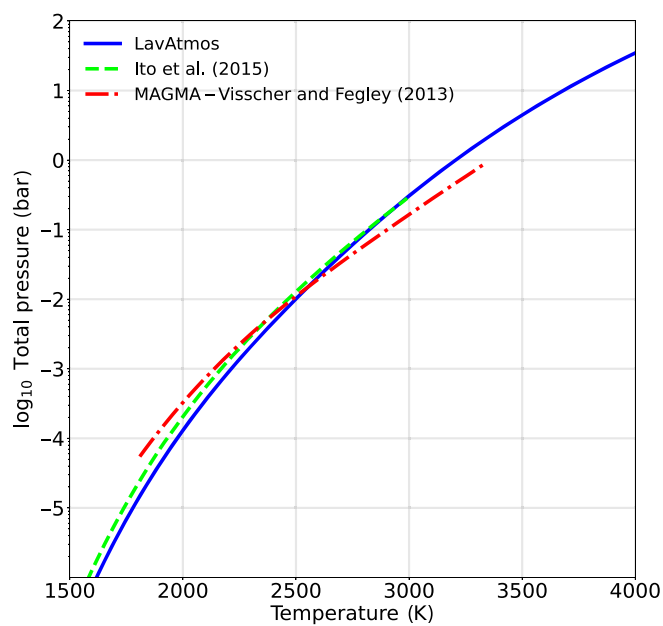

FIGURE 5. Total vapor pressures above a bulk silicate earth (BSE) melt as calculated by different codes. LavAtmos (solid dark-blue line), Ito et al. (2015) (dashed lime line), and MAGMA) (dash dotted red line) (Visscher & Fegley, 2013). (Color figure can be viewed at wileyonlinelibrary.com.)

is reached at the surface of hot rocky exoplanets (Ito et al., 2015; Miguel et al., 2011).

The value of the oxygen fugacity is derived from the law of mass balance (see "Methodology" Section). This assumes that the ratio of O atoms to cations must be the same in the out-gassed vapor as in the melt. This is a valid assumption if we assume that the system is dominated by the melt and that besides the vapor released from the lava, no other atmosphere is present. Figure 2 shows that $P_{O_2}$ values calculated using this method are in agreement with laboratory data.

The temperature range within which it is advised to use LavAtmos is between 1500 and 4000 K. For our method, we assume that the planetary surface is fully molten (above liquidus temperature). The temperature at which a surface rock composition is fully molten varies strongly with composition and pressure as shown in the work by Hirschmann (2000) who illustrate the effects on melting behavior of typical terrestrial peridotitic mantle compositions by changing the alkali content or Mg/Fe ratio of the composition. It is estimated that the liquidus of terrestrial mantle peridotite at 1 atmosphere is around 1700 K based on extrapolation of higher pressure data (Zhang & Herzberg, 1994). The liquidus temperature for the Earth's mantle at 1 atmosphere is estimated to be around 2000 K (Andrault et al., 2011). The liquidus of a basalt is down at around 1500 K (Cohen et al., 1967). Recently, it has been shown that elevated oxygen partial pressure can lower the basalt liquidus further (Lin et al., 2021). We therefore advise future users of LavAtmos to look into the estimated liquidus temperature of the composition for which to calculate degassing if a temperature below 2000 K is used. Below the liquidus temperature, crystals will form changing the composition of the magma away from the initial bulk composition. Thanks to the possibility of including mineral phases in MELTS (Ghiorso & Sack, 1995), this limit may be removed in future work, with LavAtmos incorporating information about the thermodynamic properties of the melt as it cools, and changes in melt composition due to the crystallization of minerals.

As mentioned in "Methodology" Section, the output of LavAtmos is pressure independent up to about 100 bar. Therefore, if the vaporization calculations are performed for a lava surface pressure below 100 bar, the default value of 0.1 bar can be used. If a surface pressure higher than 100 bar is expected, this should be specified to when calling LavAtmos. As mentioned before, higher pressures may change the liquidus temperature of the melt, something which should be kept in mind when selecting the temperature range for which the code is run.

The validation ("Validation" Section) with respect to laboratory data relies entirely on the data gathered by Hastie et al. (1982) and Hastie and Bonnell (1985). The first issue with this is the fact that no uncertainties were included in the published lab data. This has as consequence that the comparisons are mostly qualitative in nature. This makes it difficult to judge which of the tested codes is more representative of the experimental data. There are also only a limited number of vapor species for which partial pressure data are available. Besides the species for which data was published by Hastie and colleagues ($O_2$, K, Na, and SiO), no suitable experimental data appear to be available for the remaining species to be compared to. We are aware of experimental data on the vaporization of two lunar basalt samples (de Maria et al., 1971). Nevertheless, due to uncertainties about changes in the bulk composition of these samples during the evaporation measurements, as well as the presence of iron metal and reported interaction of oxygen with the capsule used to hold the samples (de Maria et al., 1971), we do not consider it appropriate to include these data in our comparison. This makes it difficult to judge the accuracy of the code in predicting partial pressures for which we do not have vaporization data.

Another source of uncertainty is that we are only able to compare our data to two different types of melt compositions (SRM glass and illite, see Table 2). More extensive laboratory measurements of gas partial pressures above complex melts of natural compositions are sorely needed.

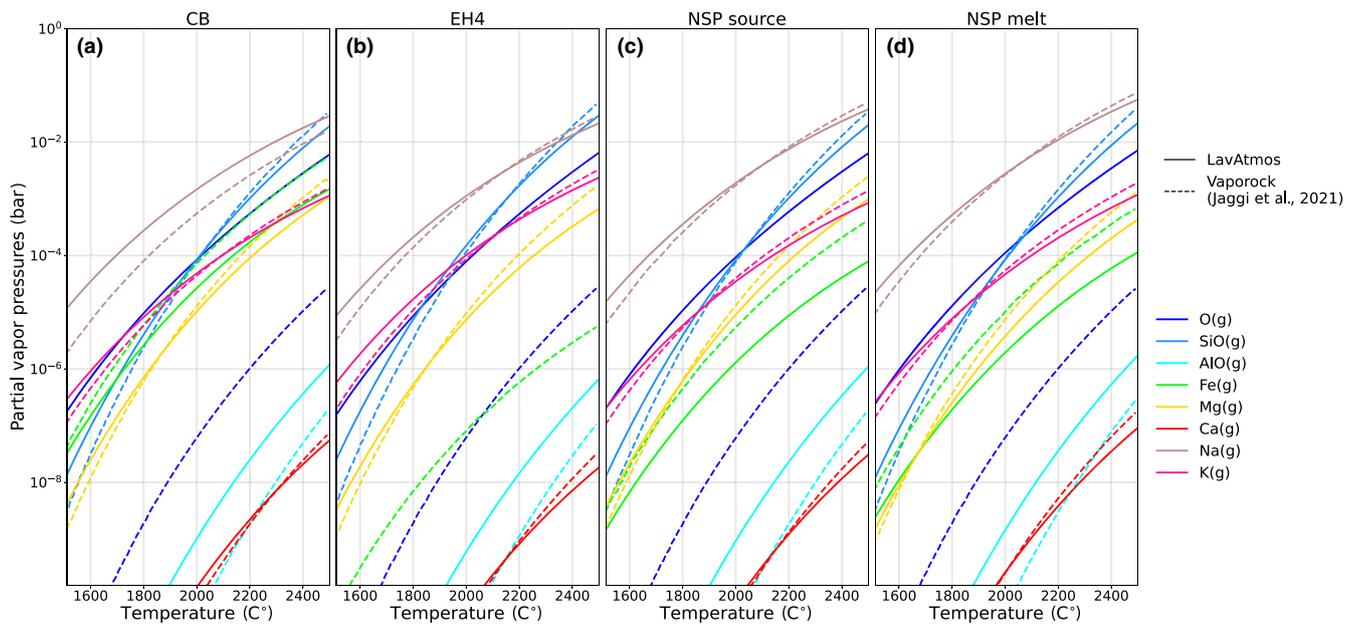

FIGURE 6. LavAtmos and VapoRock comparison. LavAtmos (solid lines) and VapoRock (dashed lines) (Wolf et al., 2023) partial pressure output for a selection of species for melt compositions CB (a), EH4 (b), Northern Smooth Planes (NSP) source (c), and NSP melt (d). See Table 2 for the compositions. (Color figure can be viewed at wileyonlinelibrary.com.)

## CONCLUSION

We developed an open-source code for gas–melt equilibrium calculations that can be used to predict the composition of a vapor above a melt of a given composition and temperature.

Thermochemical values are drawn from the JANAF tables (Chase, 1998). The geothermodynamic code MELTS (Ghiorso & Sack, 1995) is employed to model the thermodynamic activities of oxide species in the melt, which provides an approach based on more recent work than the IMCC model used by the MAGMA code. The $P_{O_2}$, necessary for determining the partial pressures of all other species, is determined self-consistently using the dual constraints of the laws of mass action and mass balance, which sets this code apart from VapoRock, which takes $P_{O_2}$ as an input. We have shown that the output of the code is in line with the available laboratory data as well as other commonly used codes from literature. LavAtmos is applicable under the condition that one can assume that chemical equilibrium is reached, that no volatiles (e.g., $H_2O$, $CO_2$, $N_2$) are included in the melt, and within a temperature range of 1500 and 4000 K. With the new generation of telescopes allowing exoplanet research to push toward observations of the atmospheres of smaller planets (JWST, Ariel, ELTs), we hope that an open-source lava-degassing code will be of use when interpreting spectra of planets potentially supporting lava oceans. Future work will involve including more volatile species so as to be applicable to a wider range of possible exoplanets.

*Acknowledgments*—This work was supported financially through a Dutch Science Foundation (NWO) Planetary and Exoplanetary Science (PEPSci) grant awarded to Yamila Miguel and Wim van Westrenen. We would like to thank Mark Ghiorso and Larry Grossman for discussions on the inner workings of the MELTS code, and the anonymous reviewers for their constructive comments on a prior version.

*Data Availability Statement*—The code presented in this work is openly available on Github at https://github.com/cvbuchem/LavAtmos.

*Editorial Handling*—Dr. Gopalan Srinivasan

## Endnotes

[1] https://enki-portal.gitlab.io/ThermoEngine/.
[2] gnu.org.
[3] The mole fraction of each of these end-member species is derived by MELTS from the melt composition given in terms of weight percentages of $SiO_2$, MgO, $Al_2O_3$, $TiO_2$, $Fe_2O_3$, FeO, CaO, $Na_2O$, and $K_2O$.
[4] https://docs.scipy.org/doc/scipy/reference/generated/scipy.optimize.fsolve.html.

[5] Activity is a term that indicates the effective concentration of a species. One can consider it as a measure of "how much" of a certain species is available for reaction.

## REFERENCES

Andrault, D., Bolfan-Casanova, N., Nigro, G. L., Bouhifd, M. A., Garbarino, G., and Mezouar, M. 2011. Solidus and Liquidus Profiles of Chondritic Mantle: Implication for Melting of the Earth across its History. *Earth and Planetary Science Letters* 304: 251–9.

Angelo, I., and Hu, R. 2017. A Case for an Atmosphere on Super-Earth 55 Cancri e. *The Astronomical Journal* 154: 232.

Asimow, P. D., and Ghiorso, M. S. 1998. Algorithmic Modifications Extending MELTS to Calculate Subsolidus Phase Relations. *American Mineralogist* 83: 1127–32.

Boukaré, C., Cowan, N. B., and Badro, J. 2022. Deep Two-Phase, Hemispherical Magma Oceans on Lava Planets. *The Astrophysical Journal* 936: 148. arXiv:2205.02864 [astro-ph, physics:physics].

Brugman, K., Phillips, M. G., and Till, C. B. 2021. Experimental Determination of Mantle Solidi and Melt Compositions for Two Likely Rocky Exoplanet Compositions. *Journal of Geophysical Research: Planets* 126: e2020JE006731. eprint. https://doi.org/10.1029/2020JE006731.

Chase, M. W. 1998. *NIST-JANAF Thermochemical Tables*, 4th ed., Vol. 1. Washington, DC: American Chemical Society.

Cohen, L. H., Ito, K., and Kennedy, G. C. 1967. Melting and Phase Relations in an Anhydrous Basalt to 40 Kilobars. *American Journal of Science* 265: 475–518.

de Maria, G., Balducci, G., Guido, M., and Piacente, V. 1971. Mass Spectrometric Investigation of the Vaporization Process of Apollo 12 Lunar Samples. *Lunar and Planetary Science Conference Proceedings*, abstract #1367, p. 2. ADS Bibcode: 1971LPSC....2.1367D.

Deibert, E. K., Mooij, E. J. W. D., Jayawardhana, R., Ridden-Harper, A., Sivanandam, S., Karjalainen, R., and Karjalainen, M. 2021. A Near-Infrared Chemical Inventory of the Atmosphere of 55 Cancri e. *The Astronomical Journal* 161: 209.

Demory, B.-O., Gillon, M., de Wit, J., Madhusudhan, N., Bolmont, E., Heng, K., Kataria, T., et al. 2016. A Map of the Large Day-Night Temperature Gradient of a Super-Earth Exoplanet. *Nature* 532: 207–9. ADS Bibcode: 2016Natur.532.207D.

Demory, B.-O., Gillon, M., Deming, D., Valencia, D., Seager, S., Benneke, B., Lovis, C., et al. 2011. Detection of a Transit of the Super-Earth 55 Cancri e with Warm *Spitzer*. *Astronomy & Astrophysics* 533: A114.

Dorn, C., and Lichtenberg, T. 2021. Hidden Water in Magma Ocean Exoplanets. *arXiv:2110.15069 [astro-ph]*.

Elkins-Tanton, L. T. 2012. Magma Oceans in the Inner Solar System. *Annual Review of Earth and Planetary Sciences* 40: 113–39.

Esteves, L. J., Mooij, E. J. W. D., Jayawardhana, R., Watson, C., and Kok, R. D. 2017. A Search for Water in a Super-Earth Atmosphere: High-Resolution Optical Spectroscopy of 55Cancri e. *The Astronomical Journal* 153: 268.

Fedkin, A., Grossman, L., and Ghiorso, M. 2006. Vapor Pressures and Evaporation Coefficients for Melts of Ferromagnesian Chondrule-like Compositions. *Geochimica et Cosmochimica Acta* 70: 206–23.

Fegley, B., and Cameron, A. 1987. A Vaporization Model for Iron/Silicate Fractionation in the Mercury Protoplanet. *Earth and Planetary Science Letters* 82: 207–22.

Ghiorso, M. S., and Gualda, G. A. R. 2015. An $H_2OCO_2$ Mixed Fluid Saturation Model Compatible with Rhyolite-MELTS. *Contributions to Mineralogy and Petrology* 169: 53.

Ghiorso, M. S., Hirschmann, M. M., Reiners, P. W., and Kress, V. C. 2002. The pMELTS: A Revision of MELTS for Improved Calculation of Phase Relations and Major Element Partitioning Related to Partial Melting of the Mantle to 3 GPa. *Geochemistry, Geophysics, Geosystems* 3: 1–35.

Ghiorso, M. S., and Sack, R. O. 1995. Chemical Mass Transfer in Magmatic Processes IV. A Revised and Internally Consistent Thermodynamic Model for the Interpolation and Extrapolation of Liquid-Solid Equilibria in Magmatic Systems at Elevated Temperatures and Pressures. *Contributions to Mineralogy and Petrology* 119: 197–212.

Greenwood, R. C., Franchi, I. A., Jambon, A., and Buchanan, P. C. 2005. Widespread Magma Oceans on Asteroidal Bodies in the Early Solar System. *Nature* 435: 916–8.

Gualda, G. A. R., Ghiorso, M. S., Lemons, R. V., and Carley, T. L. 2012. RhyoliteMELTS: A Modified Calibration of MELTS Optimized for Silica-Rich, Fluid-Bearing Magmatic Systems. *Journal of Petrology* 53: 875–90.

Hastie, J., and Bonnell, D. 1986. A Predictive Thermodynamic Model of Oxide and Halide Glass Phase Equilibria. *Journal of Non-Crystalline Solids* 84: 151–8.

Hastie, J., and Bonnell, D. W. 1985. A Predictive Phase Equilibrium Model for Multicomponent Oxide Mixtures Part II Oxides of Na-K-Ca-Mg-Al-Si. *High Temperature Science* 19: 275–306.

Hastie, J. W., Plante, E. R., and Bonnel, D. W. 1982. Alkali Vapor Transport in Coal Conversion and Combustion Systems. In *Metal Bonding and Interactions in High Temperature Systems*, Volume 179 of ACS Symposium Series, 543–600. Washington, DC: American Chemical Society, Section: 34.

Henning, W. G., Renaud, J. P., Saxena, P., Whelley, P. L., Mandell, A. M., Matsumura, S., Glaze, L. S., et al. 2018. Highly Volcanic Exoplanets, Lava Worlds, and Magma Ocean Worlds: An Emerging Class of Dynamic Exoplanets of Significant Scientific Priority. *arXiv:1804.05110 [astro-ph, physics:physics]*.

Herbort, O., Woitke, P., Helling, C., and Zerkle, A. 2020. The Atmospheres of Rocky Exoplanets: I. Outgassing of Common Rock and the Stability of Liquid Water. *Astronomy & Astrophysics* 636: A71.

Herbort, O., Woitke, P., Helling, C., and Zerkle, A. L. 2022. The Atmospheres of Rocky Exoplanets II. Influence of Surface Composition on the Diversity of Cloud Condensates. *Astronomy & Astrophysics* 658: A180. arXiv:2111.14144 [astro-ph].

Hin, R. C., Coath, C. D., Carter, P. J., Nimmo, F., Lai, Y.-J., Pogge von Strandmann, P. A. E., Willbold, M., Leinhardt, Z. M., Walter, M. J., and Elliott, T. 2017. Magnesium Isotope Evidence that Accretional Vapour Loss Shapes Planetary Compositions. *Nature* 549: 511–5.

Hirschmann, M. M. 2000. Mantle Solidus: Experimental Constraints and the Effects of Peridotite Composition. *Geochemistry, Geophysics, Geosystems* 1.

Hirschmann, M. M. 2012. Magma Ocean Influence on Early Atmosphere Mass and Composition. *Earth and Planetary Science Letters* 341-344: 48–57.

Ito, Y., Changeat, Q., Edwards, B., Al-Refaie, A., Tinetti, G., and Ikoma, M. 2021. Detectability of Rocky-Vapour Atmospheres on Super-Earths with Ariel. *Experimental Astronomy*. 53: 357–74.

Ito, Y., Ikoma, M., Kawahara, H., Nagahara, H., Kawashima, Y., and Nakamoto, T. 2015. Theoretical Emission Spectra of Atmospheres of Hot Rocky Super-Earths. *The Astrophysical Journal* 801: 144.

Jäggi, N., Gamborino, D., Bower, D. J., Sossi, P. A., Wolf, A. S., Oza, A. V., Vorburger, A., Galli, A., and Wurz, P. 2021. Evolution of Mercurys Earliest Atmosphere. *The Planetary Science Journal* 2: 230.

Keles, E., Mallonn, M., Kitzmann, D., Poppenhaeger, K., Hoeijmakers, H. J., Ilyin, I., Alexoudi, X., et al. 2022. The PEPSI Exoplanet Transit Survey (PETS) I: Investigating the Presence of a Silicate Atmosphere on the Super-Earth 55 Cnc e. *Monthly Notices of the Royal Astronomical Society* 513: 1544–56.

Kite, E. S., Fegley, B., Jr., Schaefer, L., and Ford, E. 2020. Atmosphere Origins for Exoplanet Sub-Neptunes. *The Astrophysical Journal* 891: 111. arXiv: 2001.09269.

Kite, E. S., Fegley, B., Jr., Schaefer, L., and Gaidos, E. 2016. Atmosphere Interior Exchange on Hot Rocky Exoplanets. *The Astrophysical Journal* 828: 80. arXiv: 1606.06740.

Krieger, F. J. 1965a. *The Thermodynamics of the Silica/Silicon-Oxygen Vapor System*. Technical Report. Santa Monica, CA: The Rand Corporation.

Krieger, F. J. 1965b. *The Thermodynamics of the Alumina/Aluminum Oxygen Vapor System*. Technical Report. Santa Monica, CA: The Rand Corporation.

Lamoreaux, R. H., and Hildenbrand, D. L. 1984. High Temperature Vaporization Behavior of Oxides. I. Alkali Metal Binary Oxides. *Journal of Physical and Chemical Reference Data* 13: 151–73.

Lamoreaux, R. H., Hildenbrand, D. L., and Brewer, L. 1987. HighTemperature Vaporization Behavior of Oxides II. Oxides of Be, Mg, Ca, Sr, Ba, B, Al, Ga, In, Tl, Si, Ge, Sn, Pb, Zn, Cd, and Hg. *Journal of Physical and Chemical Reference Data* 16: 419–43.

Lin, Y., van Westrenen, W., and Mao, H.-K. 2021. Oxygen Controls on Magmatism in Rocky Exoplanets. *Proceedings of the National Academy of Sciences of the United States of America* 118: e2110427118.

Miguel, Y., Kaltenegger, L., Fegley, B., and Schaefer, L. 2011. Compositions of Hot Super-Earth Atmospheres: Exploring Kepler Candidates. *The Astrophysical Journal* 742: L19.

Namur, O., Collinet, M., Charlier, B., Grove, T. L., Holtz, F., and McCammon, C. 2016. Melting Processes and Mantle Sources of Lavas on Mercury. *Earth and Planetary Science Letters* 439: 117–28.

Nguyen, T. G., Cowan, N. B., Banerjee, A., and Moores, J. E. 2020. Modelling the Atmosphere of Lava Planet K2-141b: Implications for Low- and High-Resolution Spectroscopy. *Monthly Notices of the Royal Astronomical Society* 499: 4605–12.

Nittler, L. R., and Weider, S. Z. 2019. The Surface Composition of Mercury. *Elements* 15: 33–8.

Norris, C. A., and Wood, B. J. 2017. Earths Volatile Contents Established by Melting and Vaporization. *Nature* 549: 507–10.

Palme, H., and O'Neill, H. S. C. 2003. Cosmochemical Estimates of Mantle Composition. In *Treatise on Geochemistry*, edited by H. D. Holland, and K. K. Turekian, vol. 2, 38. Elsevier Ltd: Pergamon.

Putirka, K., Dorn, C., Hinkel, N., and Unterborn, C. 2021. Compositional Diversity of Rocky Exoplanets. *arXiv:2108.08383 [astro-ph, physics:physics]*.

Putirka, K. D., and Rarick, J. C. 2019. The Composition and Mineralogy of Rocky Exoplanets: A Survey of >4000 Stars from the Hypatia Catalog. *American Mineralogist* 104: 817–29.

Schaefer, L., and Elkins-Tanton, L. T. 2018. Magma Oceans as a Critical Stage in the Tectonic Development of Rocky Planets. *Philosophical Transactions of the Royal Society A: Mathematical, Physical and Engineering Sciences* 376: 20180109.

Schaefer, L., and Fegley, B. 2007. Outgassing of Ordinary Chondritic Material and some of its Implications for the Chemistry of Asteroids, Planets, and Satellites. *Icarus* 186: 462–83.

Schaefer, L., and Fegley, B. 2009. Chemistry of Silicate Atmospheres of Evaporating Super-Earths. *The Astrophysical Journal Letters* 703: L113–7.

Schaefer, L., and Fegley, B., Jr. 2004. A Thermodynamic Model of High Temperature Lava Vaporization on Io. *Icarus* 169: 216–41.

Schaefer, L., Lodders, K., and Fegley, B. 2012. Vaporization of the Earth: Application to Exoplanet Atmospheres. *The Astrophysical Journal* 755: 41.

Sossi, P. A., and Fegley, B. 2018. Thermodynamics of Element Volatility and its Application to Planetary Processes. *Reviews in Mineralogy and Geochemistry* 84: 393–459.

Visscher, C., and Fegley, J. 2013. Chemistry of Impact-Generated Silicate MeltVapor Debris Disks. *The Astrophysical Journal* 767: L12. arXiv: 1303.3905.

Weisberg, M. K., Prinz, M., Clayton, R. N., Mayeda, T. K., Sugiura, N., Zashu, S., and Ebihara, M. 2001. A New Metal-Rich Chondrite Grouplet. *Meteoritics & Planetary Science* 36: 401–18.

Weisberg, M. K., Prinz, M., and Nehru, C. E. 1990. The Bencubbin Chondrite Breccia and its Relationship to CR Chondrites and the ALH85085 Chondrite. *Meteoritics* 25: 269–79.

Wiik, H. B. 1956. The Chemical Composition of some Stony Meteorites. *Geochimica et Cosmochimica Acta* 9: 279–89. ADS Bibcode: 1956GeCoA.9.279W.

Wolf, A. S., Jäggi, N., Sossi, P. A., and Bower, D. J. 2023. VapoRock: Thermodynamics of Vaporized Silicate Melts for Modeling Volcanic Outgassing and Magma Ocean Atmospheres. *The Astrophysical Journal* 947. https://iop science.iop.org/article/10.3847/1538-4357/acbcc7

Zhang, J., and Herzberg, C. 1994. Melting Experiments on Anhydrous Peridotite KLB-1 from 5.0 to 22.5 GPa. *Journal of Geophysical Research: Solid Earth* 99(B9): 17729–42.

Zieba, S., Zilinskas, M., Kreidberg, L., Nguyen, T. G., Miguel, Y., Cowan, N. B., Pierrehumbert, R., et al. 2022. K2 and Spitzer Phase Curves of the Rocky Ultra-Short-Period Planet K2-141 b Hint at a Tenuous Rock Vapor Atmosphere. *arXiv:2203.00370 [astro-ph]*.

Zilinskas, M., Miguel, Y., Lyu, Y., and Bax, M. 2021. Temperature Inversions on Hot Super-Earths: The Case of CN in Nitrogen-Rich Atmospheres. *Monthly Notices of the Royal Astronomical Society* 500: 2197–208. arXiv:2010.15152 [astro-ph].

Zilinskas, M., Miguel, Y., Mollire, P., and Tsai, S.-M. 2020. Atmospheric Compositions and Observability of Nitrogen

Dominated Ultra-Short Period Super-Earths. *Monthly Notices of the Royal Astronomical Society* 494: 1490–506. arXiv:2003.05354 [astro-ph].

Zilinskas, M., van Buchem, C. P. A., Miguel, Y., Louca, A., Lupu, R., Zieba, S., and van Westrenen, W. 2022. Observability of Evaporating Lava Worlds. *Astronomy & Astrophysics* 661: A126.

## APPENDIX A DERIVING THE PARTIAL PRESSURE EQUATION

For an equilibrium reaction of the type shown in Equation (1), the relation between the chemical potentials ($\mu$) of the liquid oxide species ($j$) and vapor species ($i$) involved can be written as:

$$c_{ij}\mu_j + d_{ij}\mu_{O_2} = \mu_i \tag{A1}$$

The chemical potential of a species is given by:

$$\mu = \mu^\circ + RT\ln(a) \tag{A2}$$

where $\mu^\circ$ is the standard chemical potential of the species, $a$ is the activity,[5] and $R$ the gas constant. For ideal gasses at low pressures, we assume that the activity is equal to the partial pressure of the gas. The chemical equilibrium constant for a reaction $r$ at temperature $T$ is given by:

$$\ln K_r = -\frac{\Delta_r G^\circ(T)}{RT} \tag{A3}$$

where $\Delta_f G^\circ$ is the difference in standard chemical potentials of the products and the reactants. Using this, Equation (A1) can be rewritten as:

$$c_{ij}\left(\mu_j^\circ + RT\ln(a)\right) + d_{ij}\left(\mu_{O_2}^\circ + RT\ln(P_{O_2})\right) = \mu_i^\circ + RT\ln(P_i)$$

Equation (A3) can be rewritten as:

$$-\Delta G^\circ = \mu_i^\circ - c_{ij}\mu_j^\circ - d_{ij}\mu_{O_2} = RT\ln\left(K_{r_{ij}}\right)$$

which can then be substituted into the previous equation, giving us:

$$RT\ln(P_i) = RT\ln\left(K_{r_{ij}}\right) + c_{ij}RT\ln(a_j) + d_{ij}RT\ln(P_{O_2})$$

which simplifies to Equation (3).